\documentclass[twocolumn,english,prl,showpacs,preprintnumbers,superscriptaddress]{revtex4-1}
\usepackage[utf8]{inputenc}
\usepackage{color}
\usepackage{amsmath}
\usepackage{amssymb}
\usepackage{graphicx}
\usepackage{textcomp}
\usepackage{dcolumn}
\usepackage{bm}
\usepackage[right]{eurosym}
\usepackage{float}
\usepackage[english]{babel}
\usepackage{blindtext}
\usepackage{babel}

\begin{document}

\title{A compact design for velocity-map imaging energetic electrons and ions}

\author{D. Schomas}
\affiliation{Physikalisches Institut, Universit{\"a}t Freiburg, Germany}
\author{N. Rendler}
\affiliation{Physikalisches Institut, Universit{\"a}t Freiburg, Germany}
\author{J. Krull}
\affiliation{Physikalisches Institut, Universit{\"a}t Freiburg, Germany}
\author{R. Richter}
\affiliation{Elettra Syncrotrone, Trieste, Italy}
\author{M. Mudrich}
\email{mudrich@physik.uni-freiburg.de}
\affiliation{Department of Physics and Astronomy, Aarhus University, Denmark}

\begin{abstract}
We present a compact design for a velocity-map imaging spectrometer for energetic electrons and ions. The standard geometry by Eppink and Parker [A. T. J. B. Eppink and D. H. Parker, Rev. Sci. Instrum. \textbf{68}, 3477 (1997)] is augmented by just two extended electrodes so as to realize an additional einzel lens. In this way, for a maximum electrode voltage of $7~$kV we experimentally demonstrate imaging of electrons with energies up to 65~eV. Simulations show that energy acceptances of $\lesssim 270$ and $\lesssim 1,200$~eV with an energy resolution of $\Delta E / E \lesssim 5\,$\% are achievable for electrode voltages of $\le 20$~kV when using diameters of the position-sensitive detector of 42 and 78~mm, respectively. 
\end{abstract}

\date{\today}

\maketitle
\section{Introduction}
Velocity-map imaging (VMI) of electrons and ions is an established technique in the fields of atomic, molecular and chemical physics~\cite{Whitaker}. The advantages of electron and/or ion VMI compared to other detection techniques are the high collection efficiency and the high level of detail at which information can be obtained about the dynamics of the photoionization process, including ion fragment masses and velocities, photoelectron energies, as well as ion and electron angular distributions. While still mostly used in laser-based experiments, the VMI technique is becoming increasingly popular for use with synchrotron radiation~\cite{Garcia:2005,Keeffe:2011}, free-electron lasers~\cite{Rouzee:2011,Lyamayev:2013,Fukuzawa:2016}, and laser-based sources of extreme ultraviolet (XUV) radiation~\cite{Kornilov:2010,Schuette:2014,Kling:2014}. Besides, the interaction of intense laser pulses with atoms, molecules and clusters leading to strong-field ionization and the ignition of a nanoplasma in nanoclusters is being studied using VMI spectrometers~\cite{Schuette:2015}. Both high photon energies reaching up to the XUV and even into the X-ray spectral regions, as well as strong-field ionization processes, demand techniques to detect electrons and ions with high kinetic energies up to the keV-range. 

In principle, the energy acceptance of VMI spectrometers can be extended by (i) increasing the size of the position-sensitive detector, (ii) reducing the flight distance of the charged particles from the interaction region up to the detector, (iii) increasing the acceleration voltages applied to electrodes, and implementing appropriate design changes pertaining to insulators, feedthroughs, etc., or (iv) by adding ion optical elements to refocus the charged particles onto the detector. 

To detect higher kinetic energy particles, the traditional design of Eppink and Parker~\cite{Eppink:1997}, which consists of one filled repeller plate electrode and two open extractor electrodes, was modified by Garcia~\textit{et al.}~\cite{Garcia:2005} who applied modification (iv). Similarly to the original design where an einzel lens was added that magnifies the images of energetic electrons or ions~\cite{Offerhaus:2001}, Garcia~\textit{et al.} introduced a double einzel lens in the flight tube to refocus electrons onto the detector~\cite{Garcia:2005}. For a detector with an active area of 36~mm in diameter, the authors demonstrated electron energies up to $E=14~$eV and a relative energy resolution $\Delta E/E=6~$\%. Energy acceptances up to several hundred eV were predicted for the case of additionally applying modification (iii). 

In 2014, two new designs were presented which are explicitly optimized for imaging energetic charged particles. The design by Skruszewicz~\textit{et al.}~\cite{Skruszewicz:2014} combines modifications (i)-(iv). In particular, by inserting two additional electrodes set to opposite polarity with respect to the repeller and extractor electrodes into a shortened version of the traditional VMI geometry, the authors achieved a high energy acceptance $\lesssim 500~$eV. This approach has the advantage of slightly reducing the curvature of the focal plane despite the short flight distance, similar to an achromatic lens. Kling~\textit{et al.} presented a thick-lens design consisting of 11 electrodes all set to equal polarity and implementing modifications (i)-(iv)~\cite{Kling:2014}. The authors demonstrate high energy resolution $\Delta E/E\lesssim 4~$\% for electrons up to $\sim 1,000~$eV. Naturally, such a complex electrode arrangement complicates the construction and operation of the spectrometer, especially when applied to high-energy particle imaging. Most recently, Fukuzawa~\textit{et al.} were able to detect electrons up to 960~eV, using a double-sided VMI with traditional electrode geometry only by implementing modifications (i)-(iii). However, impractically high electrode voltages up to 27~kV were used~\cite{Fukuzawa:2016}. Such high voltages require more specialized connectors, feedthroughs, electrode and insulator materials and surface finish, which significantly impacts the compactnes and costs of a VMI spectrometer. 

The spectrometer design presented here is intended for VMI studies of nanoplasmas created by strong-field ionization of rare gas clusters and helium (He) nanodroplets~\cite{KrishnanPRL:2011,Krishnan:2012,Heidenreich:2016,Heidenreich:2017}. It fulfills the requirement of providing a wide gap between repeller and extractor electrodes to allow for placing a curved mirror near the interaction region for tight focusing of the laser beam, while extending the energy acceptance to the keV-range. The design resorts to adding a refocusing lens to the traditional three-plates geometry in addition to implementing modifications (ii) and (iii). 
In contrast to the design by Garcia~\textit{et al.}~\cite{Garcia:2005}, the additional ion optics consists of just two extended electrodes added to the traditional three-plate geometry. We find this design to be the best compromise between modifications (ii) and (iii) for achieving a wide energy acceptance and reaching sufficiently high energy resolution, while keeping the complexity of the electrode arrangement and the associated technical effort and costs within a limit. We systematically characterize the performance of our design by carrying out trajectory simulations. Test measurements using a diode laser and synchrotron radiation to create electrons with energies up to 65~eV are in good agreement with the simulations.

\section{\label{sec:Design}Spectrometer design}
The spectrometer design is inspired by the one of Garcia~\textit{et al.}~\cite{Garcia:2005} (additional einzel lens) as well as the one of Skruszewicz~\textit{et al.}~\cite{Skruszewicz:2014} (additional achromatic lens realized by two plate electrodes). It is the result of systematic simulations of electron trajectories using the SIMION8.0 package to assess the maximum energy acceptance and optimum energy resolution. A set of 27 electrons distributed regularly over a cubic volume of $0.6\times 0.6\times 0.6$~mm$^3$ was used as an input into the simulation. For each value of the electron energy, emission angles were varied from $10^\circ$ to $170^\circ$ in steps of $10^\circ$. For a fixed geometry of the repeller and extractor electrodes, the addition of a complete einzel lens consisting of 3 extended ring electrodes with different values of the diameter and length was considered in the simulations. Both cylindrical and conical electrodes were tested. The same procedure was repeated for a lens system consisting of 2 electrodes. The latter was found to be equally performant in terms of resolution and accepance. Therefore, we chose the simpler geometry consisting of only 2 lens electrodes.

\begin{figure}
\centering
\includegraphics[width=0.45\textwidth]{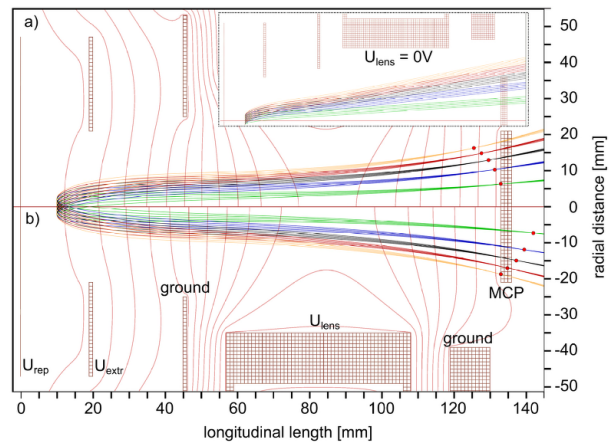}
\caption{\label{fig:trajectories} Cross sectional view of the electrode arrangement and electron trajectories for various initial positions and transversal velocities. The inset shows the trajectories for the voltage applied to the lens electrode which is set to zero potential.}
\end{figure}
Fig.~\ref{fig:trajectories} shows a cross-sectional view of the electrode geometry of our spectrometer as well as electron trajectories for various initial positions along a line of length 5~mm and kinetic energies ranging between 5.4 and 45.4~eV. The electrode voltages are set to $U_\mathrm{rep}=-3,000$~V at the repeller and, $U_\mathrm{ext}=-2,320$~V at the extractor. The voltage applied to the extended lens electrode is $U_\mathrm{lens}= +5,240$~V and $+4,685$~V in Fig.~\ref{fig:trajectories} a) and b), respectively. The front side of the MCP is set to ground potential whereas the back side is set to $+1,500$~V. The remaining two electrodes are set to ground potential. The inset shows electron trajectories for the same initial conditions, electrode geometry, and voltages except that $U_\mathrm{lens}=0$. 

In our compact design, the two additional extended electrodes in combination with the second grounded extractor electrode of the traditional geometry together act as an einzel lens. Equipotential lines, depicted as thin red lines in Fig.~\ref{fig:trajectories}, follow the typical pattern of an einzel lens in that region of the spectrometer. We find this to be a simple, yet very efficient way of achieving refocusing conditions for energetic charged particles. For detection we use a micro-channel plate (MCP) detector with an active area of 42~mm in combination with a phosphor screen. In contrast to other VMI configurations, we choose the extractor-to-repeller voltage ratio such that the velocity focus with refocusing lens switched off ($U_\mathrm{lens}=0$) is located at infinite distance. Thus, electrons emitted from different points of origin along a vertical line but with identical kinetic energy, fly along nearly parallel trajectories, see inset of Fig.~\ref{fig:trajectories}. The additional lens system focuses these parallel trajectories onto the detector plane. 
This allows us to vary the focus position and the points of incidence of electrons onto the detector plane nearly independently. 

In Fig.~\ref{fig:trajectories} the focus positions are depicted by red dots; in the upper half, where the lens voltage is set to $U_\mathrm{lens}=+5,240$~V, the lowest-energetic electrons are focused onto the detector and for higher energetic ones the foci lie in front of the detector. In the lower half the lens voltage is set to $U_\mathrm{lens}=+4,685$~V such that the highest-energy electrons are focused onto the detector and the foci for slower electrons fall behind the detector plane. Thus, the energy resolution is dependent on the electron energy, known as chromatic aberration. For the voltage setting in Fig.~\ref{fig:trajectories} a) we obtain $\Delta E/E = 11\,$\% for low electron energies ($E=5.4$~eV) and $3.9\,$\% for high energies (45~eV). Note that these simulations are carried out by letting the electrons fly isotropically out of the initial volume, in contrast to the trajectories shown in Fig.~\ref{fig:trajectories}. These values only slightly differ from those obtained for the voltages used in Fig.~\ref{fig:trajectories} b), $\Delta E/E = 16\,$\% for $E=5.4$~eV and $3.4\,$\% for 45~eV. However, the points of incidence on the detector slightly differ for the two voltage settings, corresponding to an energy shift of 8\,\% of the measured value of $E$ for a given energy calibration. Thus, if various focusing conditions are to be used in one experiment, a calibration measurement should be recorded for each voltage setting.

Besides substantially increasing the energy acceptance for energetic charged particles by applying a high positive voltage $U_\mathrm{lens}$, our design provides further advantages. 
The short drift distance makes the setup less sensitive to stray magnetic fields. For this reason, we only use a single unbaked $\mu$-metal shield with holes big enough for a laser beam of 25~mm in diameter to enter into the interaction region. Furthermore, the flight region from the extractor up to the detector is well shielded by the extended lens electrodes against stray electric fields that may originate from cables or patch charges on insulator surfaces.

\begin{figure}
\centering
\includegraphics[width=0.45\textwidth]{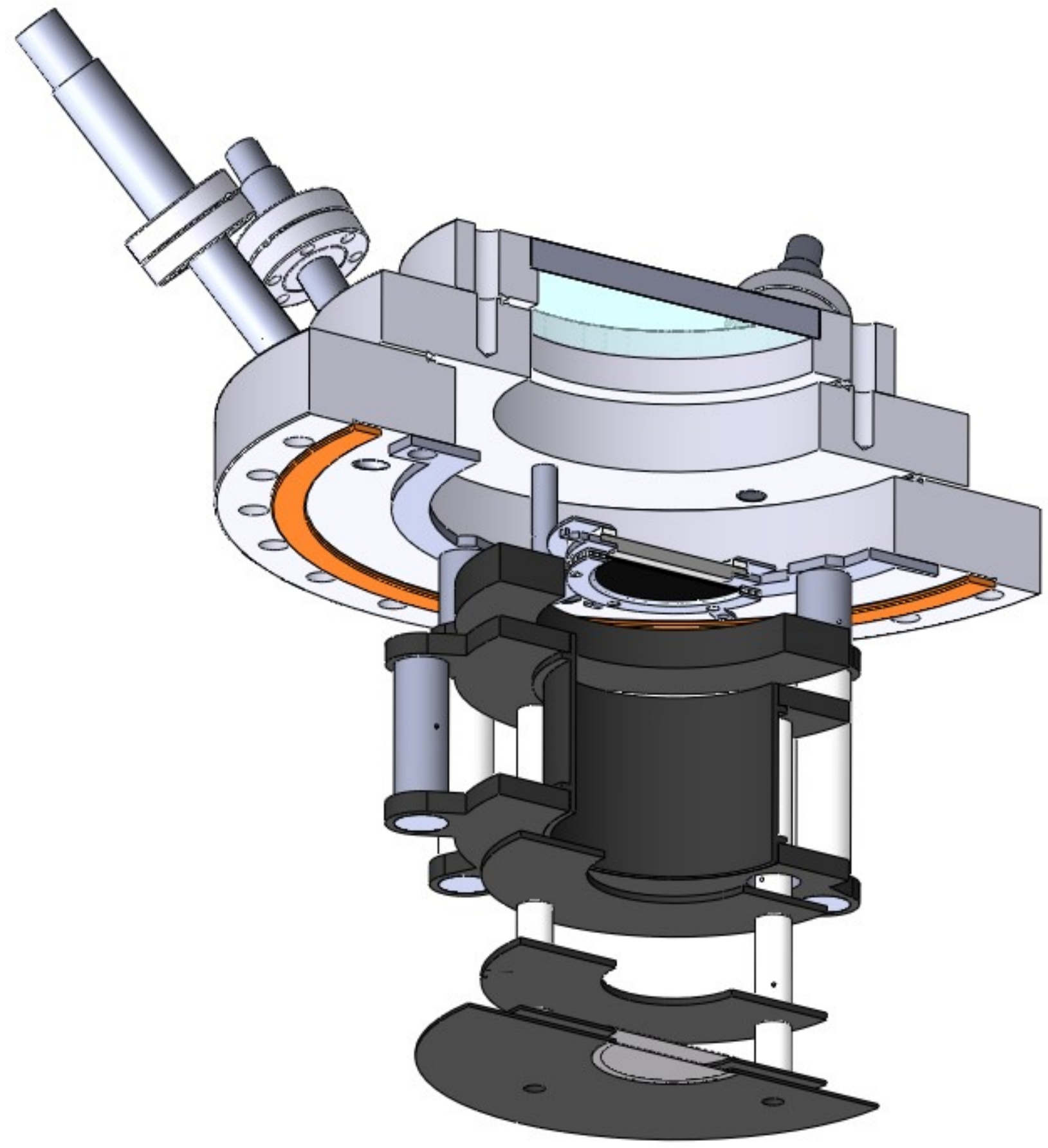}
\caption{\label{fig:setup} Cross sectional view of the compact VMI setup. Dark areas indicate metal electrodes, white parts are PEEK insulators.}
\end{figure}
The technical realization of our compact VMI setup is shown in Fig.~\ref{fig:setup}. The entire electrode system is mounted on a CF160-to-CF200 zero-length adaptor flange. High-voltage coaxial feedthroughs (SHV rated to 20~kV) mounted in CF16 flanges are connected to half-nipples welded into the adaptor flange. The imaging detector consists of two MCPs in Chevron configuration with an active area of 42~mm diameter (TOPAG Lasertechnik GmbH) and a phosphor screen (ProxiVision GmbH). It is separately mounted to a CF100-to-CF160 zero-length adaptor flange. This allows us to easily replace the detector by a larger one, or by a delay-line detector. A non-magnetic CF100 window flange closes the stack of three flanges. 

The stack of electrodes (black parts in Fig.~\ref{fig:setup}) consists of two closely-spaced plates forming the repeller, each with central holes covered by copper meshes with high transmission, two open extractor electrodes, one cylindrical lens electrode, and one thick ground electrode in front of the detector. The repeller electodes are designed such that the setup can be extended by an additional time-of-flight spectrometer for ions, while minimizing punch-through electric fields. The diameters of the central holes in the electrodes are 40~mm (extractor 1), 48~mm (extractor 2), 68~mm (lens), 76~mm (ground). The length of the lens electrode is 52~mm, that of the last ground electrode is 11~mm. The total distance between the repeller and detector surfaces amounts to 136~mm. The distance between repeller and extractor electrodes is 25~mm to leave enough space for having the option of placing a focusing mirror near the interaction volume for tightly focusing a laser beam of up to 1'' in diameter. All electrodes are made of aluminum, which we coat with graphite using an air brush. Each electrode is mounted by threaded polyetheretherketone (PEEK) cylinders (white parts in Fig.~\ref{fig:setup}). Due to the tendency of PEEK surfaces to charge up electrostatically, we noticed that graphite particles can be attracted from the electrode surfaces onto the insulators, thereby reducing the withstand voltage. Therefore, we recommend to use ceramic insulators if graphitizing of the electrode surfaces is indispensable. 

\begin{figure}
\centering
\includegraphics[width=0.4\textwidth]{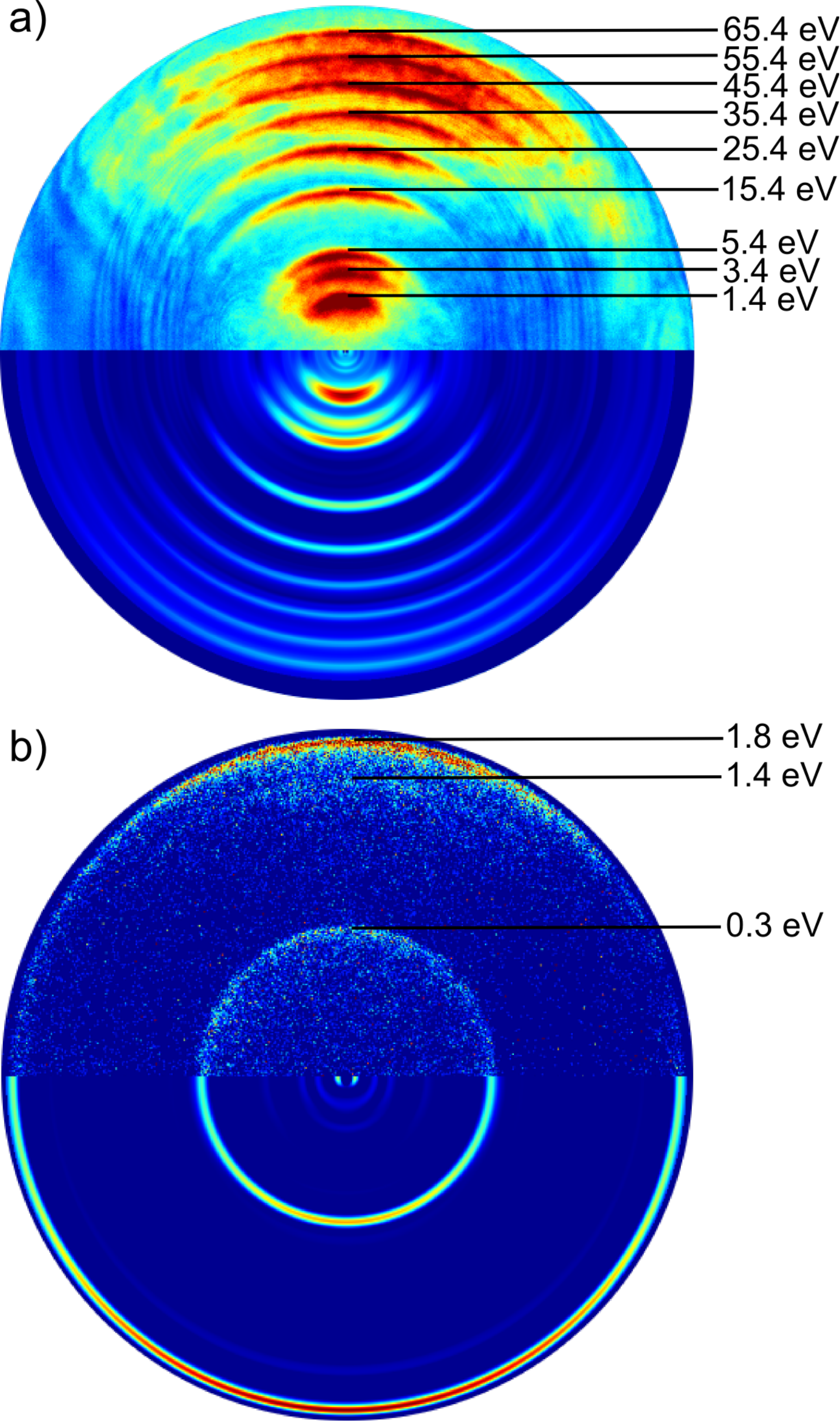}
\caption{\label{fig:rawimages} a) Sum of 8 velocity map electron images recorded by photoionizing helium atoms at photon energies 26, 28, 30, 40, 50, 60, 70, 80, 90~eV. b) Electron image recorded by resonantly photoionizing potassium atoms using a diode laser. The upper half of each image shows the raw data, the lower half shows the result of Abel inversion.}
\end{figure}
\section{\label{sec:Results}Test measurements}
To test our spectrometer design we have performed two experiments in electron VMI detection mode. The first experiment was carried out at the Gasphase beamline of Elettra Syncrotrone, Trieste, Italy. There, a continuous supersonic beam of He atoms was generated by expanding pressurized He gas at room temperature out of a 5~$\mu$m nozzle into vacuum. The XUV beam intersected the He beam at right angles inside the VMI spectrometer. The interaction volume in this case is estimated to be of the size $0.5\times 0.5\times 2~$mm$^3$. For these measurements we chose a fixed voltage setting ($U_\mathrm{rep}=-4,980~$V, $U_\mathrm{ext}=-3,820~$V, $U_\mathrm{lens}=+7,000~$V) and scanned the photon energy from $25~$eV up to $90~$eV. This voltage setting is adjusted to obtain highest resolution for the high-energy electrons ($E=65.4$~eV). The reason for not using higher voltages $U_\mathrm{lens}$ to further increase the energy acceptance was that sparking occured along the PEEK insulators caused by deposits of graphite on the insulator surfaces originating from the electrode coatings. Note that after removing all graphite coatings at a later stage, $U_\mathrm{lens}=+20~$kV was reached without any signs of discharging, showing that the mechanical design is in principle well-suited for the targeted high voltages. 

The raw photoelectron images are background subtracted, summed up, and the result is displayed in the upper half of Fig.~\ref{fig:rawimages}. 
The ring-like structures result from the projections of the directly emitted photoelectrons onto the detector plane. The anisotroply of the angular distribution reflects the probability distribution of a one-photon transition out of the He groundstate into the ionization continuum induced by the synchrotron radiation, which is linearly polarized perpendicularly to the spectrometer axis. The relative resolution is determined from the individual images by performing inverse Abel transformation (see Fig.~\ref{fig:rawimages}, lower halves of images), integration over angles, and fitting the peaks with gaussian functions. 

\begin{figure}
\centering
\includegraphics[width=0.5\textwidth]{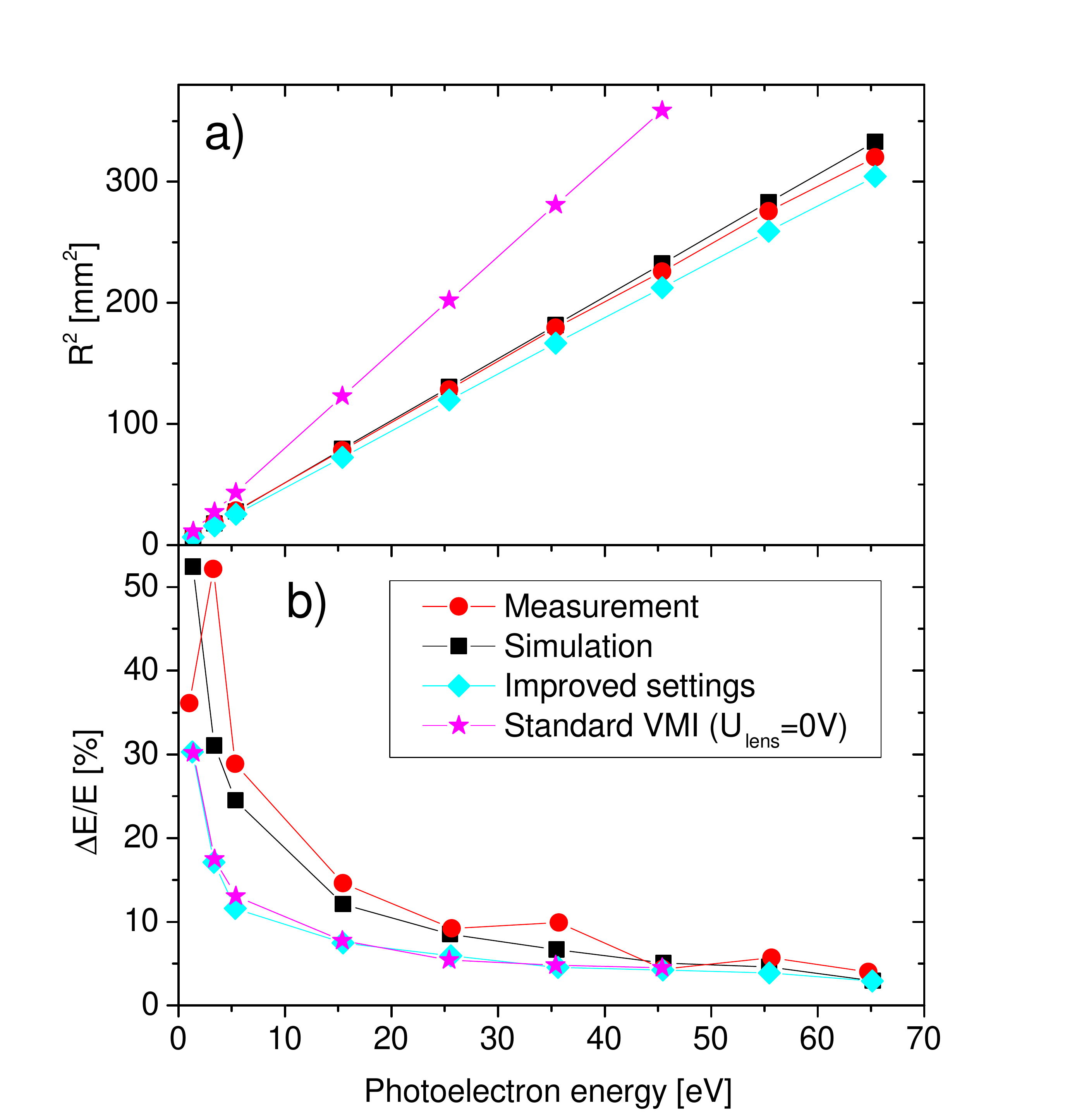}
\caption{\label{fig:resolution} Comparison of the simulated and experimental characteristics of the spectrometer. a) Radius of the electron signal in the detector plane for various settings of high voltages; b) Relative energy resolution.}
\end{figure}
Fig.~\ref{fig:resolution} a) shows the square of the fitted peak positions as a function of the calculated photoelectron energy $E=h\nu - E_I$ as red dots, where $h\nu$ is the photon energy and $E_I=24.6~$eV denotes the ionization energy of He. Thus, we experimentally demonstrate electron imaging with high resolution up to a maximum electron energy of 65~eV, only limited to the mentioned technical issue. The red filled circles in Fig.~\ref{fig:resolution} b) represent the relative energy resolution $\Delta E/E$ where we define $\Delta E$ as the full width at half maximum (FWHM) of the fitted peak function. 

The black squares show the results of our trajectory simulations for the same voltage setting as in the experiment. In the simulation, the distribution of initial positions was assumed to be of gaussian shape in three dimensions with standard deviations of 1~mm along the spectrometer axis and $1~$mm$\times 1.5$~mm in the perpendicular directions. The simulation was carried out for $2\times 10^4$ to $3\times 10^5$ electrons from the lowest to the highest energies, respectively. The electron velocities were distributed isotropically. The simulations are in good agreement with the measurements except for the low energy range ($E=1.4$ and $3.4~$eV), where weak distortions of the electric field become apparent.

Unfortunately, the voltages used in the experiment were imperfect, which can be seen by comparing the results with those from the simulation for voltages at which electrons are optimally focused on the detector for each electron energy (light blue diamonds). The corresponding voltages are $U_\mathrm{rep} = -4,980~$kV, $U_\mathrm{ext} = -3,884~$V, and $U_\mathrm{lens} = 8,964~$V. The reason for not setting the optimum voltages was again our limitation to a maximum $U_\mathrm{lens}=7~$kV. 

For comparison, the results from simulations with standard VMI conditions~\cite{Eppink:1997} where $U_\mathrm{rep} = -4,980~$kV is held constant but $U_\mathrm{lens}=0$, are shown in Fig.~\ref{fig:resolution} as pink stars. While the resolution is nearly identical to the one for optimized setting in the refocusing mode of operation [diamonds in Fig.~\ref{fig:resolution} b)], the peak positions shown in a) are shifted to larger radii and thus the energy acceptance is greatly reduced.

A second experiment was performed to assess our design at low electron kinetic energies in the range of a few eV. To this end we photoionized potassium atoms generated in an effusive beam out of a heated crucible using a frequency-stabilized diode laser according to the scheme described in Ref.~\cite{Wituschek:2016}. For these measurements we used the voltage setting $U_\mathrm{rep}= -600 ~$V, $U_\mathrm{ext}=-480 ~$V, and $U_\mathrm{lens}=+850~$V. A typical raw image taken with an exposure time of a few seconds is displayed in Fig.~\ref{fig:rawimages} b). It shows three rings due to photoionization of excited potassium atoms out of three different electronic states. The resulting photoelectron energies are $E=0.34~$eV measured with a resolution of 11\,\%, $1.4$~eV (7.6\,\% resolution), and $1.8~$eV (14\,\% resolution). These results clearly demonstrate that our VMI design is able to image both low and high-energy electrons with good resolution even when using a detector with only 42~mm diameter of the active area.

\begin{figure}
\centering
\includegraphics[width=0.5\textwidth]{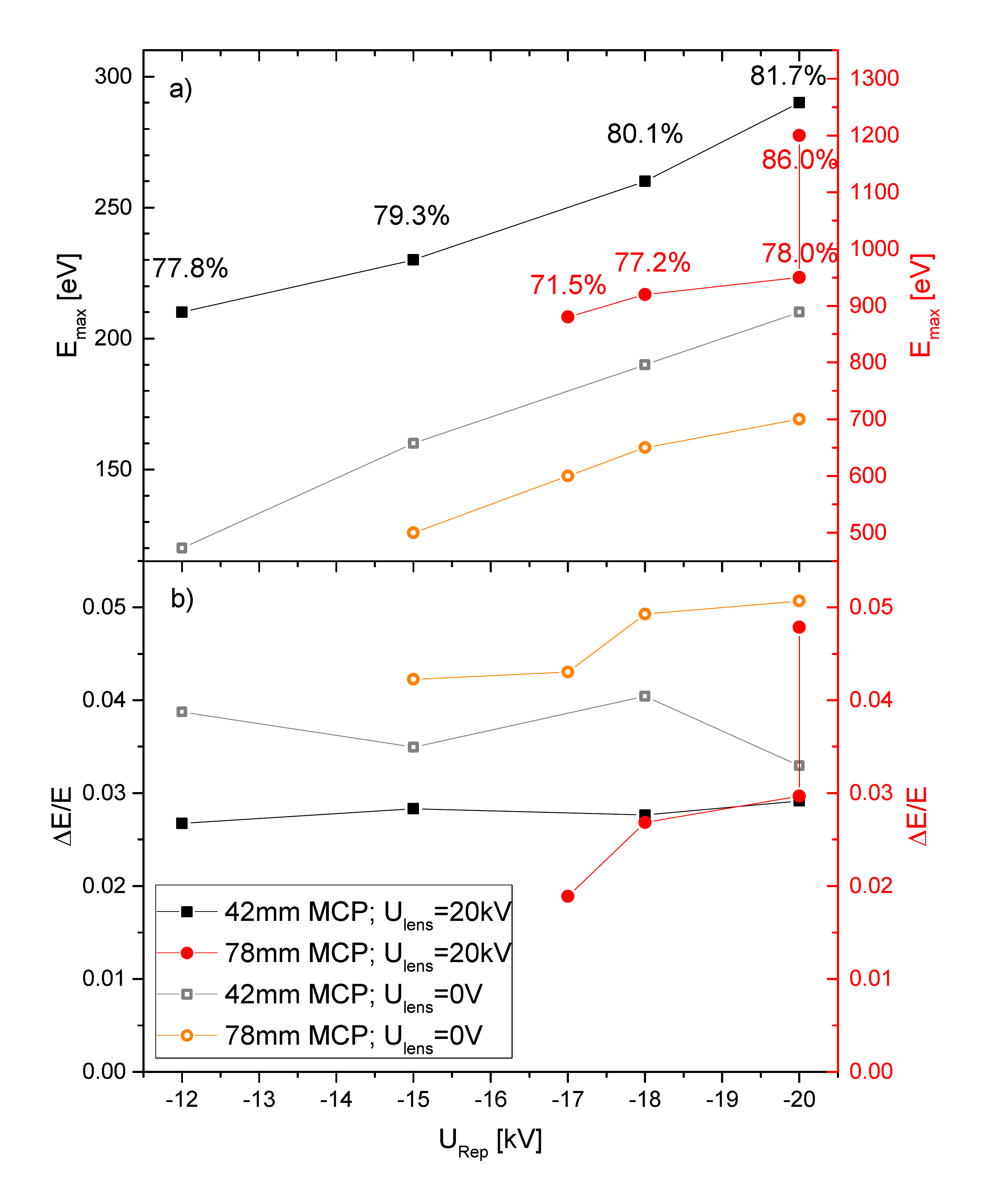}
\caption{\label{fig:extended} a) Simulated energy acceptance $E_\mathrm{max}$ and b) relative resolution $\Delta E/E$ of the spectrometer as a function of voltage applied to the repeller electrode $U_\mathrm{rep}$. For each data point the ratio of extractor and repeller voltages is adjusted as indicated. The lens voltage $U_\mathrm{lens}$ is set to 0 (open symbols) or 20~kV (closed symbols). Data points for the 42~mm MCP detector (squares) are referenced to the $E_\mathrm{max}$-scale shown on the left-hand side, for the 78~mm MCP detector (circles) the scale on the right-hand side applies.} 
\end{figure}
Finally, we systematically assess the performance of our spectrometer design and explore the potential to further push the acceptance to higher energies when allowing for slightly inferior energy resolution. Fig.~\ref{fig:extended} a) shows the simulated maximum detectable electron kinetic energies $E_\mathrm{max}$ for MCP detectors with 42 and 78~mm diameter. The corresponding relative energy resolution $\Delta E/E$ in shown in Fig.~\ref{fig:extended} b). Filled symbols represent the results for the maximum design value of the lens voltage, $U_\mathrm{lens}=20~$kV. Open symbols show the results for $U_\mathrm{lens}=0$ for comparison. For $U_\mathrm{lens}=0$, the optimum ratio between extractor and repeller voltages in terms of optimum resolution was $U_\mathrm{ext}/U_\mathrm{rep}= 88\,$\% for the MCP with a diameter of  42~mm and $U_\mathrm{ext}/U_\mathrm{rep}= 84\,$\% for the 78~mm MCP.

For the 42~mm MCP, the optimum setting for the repeller and extractor voltages at $U_\mathrm{lens}=20~$kV are $U_\mathrm{rep}=-12~$kV and $U_\mathrm{ext}=-9.3~$kV, respectively. The corresponding voltage ratio $U_\mathrm{ext}/U_\mathrm{rep}= 78\,$\% is indicated in Fig.~\ref{fig:extended} a) as a label next to the data point. The resulting value $E_\mathrm{max}=210~$eV matches the value $E_\mathrm{max}=210~$eV obtained when the einzel lens is not used ($U_\mathrm{lens}=0$), but instead a maximum voltage $U_\mathrm{rep}=-20~$kV is applied to the repeller. However, the energy resolution $\Delta E/E=2.7\,$\% is slightly better than for the reference setting ($U_\mathrm{lens}=0$, $U_\mathrm{rep}=-20~$kV) for which $\Delta E/E=3.5\,$\%. For the 78~mm MCP and when the lens is used ($U_\mathrm{lens}=20~$kV), the optimum voltage setting is $U_\mathrm{rep}=-17~$kV and $U_\mathrm{ext}=-12.2~$kV ($72\,\%$). In this case, the resulting energy acceptance $E_\mathrm{max}=880~$eV significantly exceeds the value $E_\mathrm{max}=700~$eV obtained for the reference setting, and the resolution is notably improved ($\Delta E/E=1.9\,$\% versus $5.0\,$\% for $U_\mathrm{lens}=0$). 

Thus, when using the 42~mm MCP, the lens does not significantly improve the performance of the spectrometer for a given maximum voltage applied to either lens or repeller electrodes, if optimum resolution is required. When using the larger 78~mm MCP, however, the lens substantially enhances both $E_\mathrm{max}$ and $\Delta E/E$. This is owing to the stronger refocusing effect of the lens for energetic electrons, which keeps the trajectories at a distance from fringe fields created by the edges of electrodes, and thus reduces aberations. 

Furthermore, if a slightly lowered energy resolution is tolarable, using the lens at $U_\mathrm{lens}=20~$kV raises the energy acceptance for the 42~mm MCP to $E_\mathrm{max}=270~$eV if setting $U_\mathrm{rep}=-20~$kV and $U_\mathrm{ext}/U_\mathrm{rep}=82\,$\%. The energy resolution then slightly deterorates to 3.0\,\%, which may still be acceptable for experiments using broadband radiation or strong-field ionization. When using a 78~mm MCP the maximum detectable kinetic energy can even be increased to the keV-range if an energy resolution $\lesssim 5.0\,$\% is tolerable.

\section{Summary}
To summarize, we present a compact design for a VMI spectrometer optimized for detecting energetic charged particles generated, e.\,g., in a laser-induced nanoplasma. The main features are (i) a simple construction, which extends the traditional three-plate geometry~\cite{Eppink:1997} by just two additional cylindrical electrodes. This allows for a modular mounting of the electrode system and the MCP detector on two stacked vacuum flanges; (ii) a wide gap between repeller and extractor electrodes which allows for good access e.\,g. for placing a focusing mirror near the interaction region; (iii) the compatibility with high voltages up to 20~kV. In combination with a short flight distance and an added einzel lens we achieve an enhanced energy acceptance and an improved energy resolution, in particular when using a large imaging detector (78~mm diameter). 

While we experimentally demonstrate electron imaging up to an electron energy of 65~eV with a resolution $<4\,$\% using a 42~mm detector, limited only by technical issues, simulations show that maximum energies up to 270~eV are achievable when tolerating a slightly diminished energy resolution $\lesssim 3\,$\%. We find that our design compares favorably with respect to previously reported designs~\cite{Kling:2014,Skruszewicz:2014} in terms of energy acceptance and resolution while being particularly simple and compact. 
In a future upgrade the MCP detector will be replaced by one with a diameter of 78~mm, which will further increase the energy acceptance to 850~eV at high resolution $\lesssim 2\,$\% and up to 1,200~eV if allowing for a slightly inferior resolution $\lesssim 5\,$\%. In the planned experiments on nanoplasmas induced by broad-band ultrahort laser pulses the latter resolution will be by far sufficient whereas the ability of detecting high energy electrons and ions will be crucial.

\begin{acknowledgments}
The authors gratefully acknowledge fruitful discussions with J. von Vangerow as well as financial support by the Deutsche Forschungsgemeinschaft within projects MU 2347/10-1 and MU 2347/12-1 in the frame of the Priority Programme 1840 `Quantum Dynamics in Tailored Intense Fields'.
\end{acknowledgments}


\begin{thebibliography}{18}%
	\makeatletter
	\providecommand \@ifxundefined [1]{%
		\@ifx{#1\undefined}
	}%
	\providecommand \@ifnum [1]{%
		\ifnum #1\expandafter \@firstoftwo
		\else \expandafter \@secondoftwo
		\fi
	}%
	\providecommand \@ifx [1]{%
		\ifx #1\expandafter \@firstoftwo
		\else \expandafter \@secondoftwo
		\fi
	}%
	\providecommand \natexlab [1]{#1}%
	\providecommand \enquote  [1]{``#1''}%
	\providecommand \bibnamefont  [1]{#1}%
	\providecommand \bibfnamefont [1]{#1}%
	\providecommand \citenamefont [1]{#1}%
	\providecommand \href@noop [0]{\@secondoftwo}%
	\providecommand \href [0]{\begingroup \@sanitize@url \@href}%
	\providecommand \@href[1]{\@@startlink{#1}\@@href}%
	\providecommand \@@href[1]{\endgroup#1\@@endlink}%
	\providecommand \@sanitize@url [0]{\catcode `\\12\catcode `\$12\catcode
		`\&12\catcode `\#12\catcode `\^12\catcode `\_12\catcode `\%12\relax}%
	\providecommand \@@startlink[1]{}%
	\providecommand \@@endlink[0]{}%
	\providecommand \url  [0]{\begingroup\@sanitize@url \@url }%
	\providecommand \@url [1]{\endgroup\@href {#1}{\urlprefix }}%
	\providecommand \urlprefix  [0]{URL }%
	\providecommand \Eprint [0]{\href }%
	\providecommand \doibase [0]{http://dx.doi.org/}%
	\providecommand \selectlanguage [0]{\@gobble}%
	\providecommand \bibinfo  [0]{\@secondoftwo}%
	\providecommand \bibfield  [0]{\@secondoftwo}%
	\providecommand \translation [1]{[#1]}%
	\providecommand \BibitemOpen [0]{}%
	\providecommand \bibitemStop [0]{}%
	\providecommand \bibitemNoStop [0]{.\EOS\space}%
	\providecommand \EOS [0]{\spacefactor3000\relax}%
	\providecommand \BibitemShut  [1]{\csname bibitem#1\endcsname}%
	\let\auto@bib@innerbib\@empty
	\bibitem [{\citenamefont {Whitaker}(2003)}]{Whitaker}%
	\BibitemOpen
	\bibfield  {author} {\bibinfo {author} {\bibfnamefont {B.~J.}\ \bibnamefont
			{Whitaker}},\ }\href@noop {} {\emph {\bibinfo {title} {Imaging in Molecular
				Dynamics}}}\ (\bibinfo  {publisher} {Cambridge University Press},\ \bibinfo
	{year} {2003})\ p.\ \bibinfo {pages} {249}\BibitemShut {NoStop}%
	\bibitem [{\citenamefont {Garcia}\ \emph {et~al.}(2005)\citenamefont {Garcia},
		\citenamefont {Nahon}, \citenamefont {Harding}, \citenamefont {Mikajlo},\
		and\ \citenamefont {Powis}}]{Garcia:2005}%
	\BibitemOpen
	\bibfield  {author} {\bibinfo {author} {\bibfnamefont {G.~A.}\ \bibnamefont
			{Garcia}}, \bibinfo {author} {\bibfnamefont {L.}~\bibnamefont {Nahon}},
		\bibinfo {author} {\bibfnamefont {C.~J.}\ \bibnamefont {Harding}}, \bibinfo
		{author} {\bibfnamefont {E.~A.}\ \bibnamefont {Mikajlo}}, \ and\ \bibinfo
		{author} {\bibfnamefont {I.}~\bibnamefont {Powis}},\ }\href@noop {}
	{\bibfield  {journal} {\bibinfo  {journal} {Rev. Sci. Instrum.}\ }\textbf
		{\bibinfo {volume} {76}},\ \bibinfo {pages} {053302} (\bibinfo {year}
		{2005})}\BibitemShut {NoStop}%
	\bibitem [{\citenamefont {O’Keeffe}\ \emph {et~al.}(2011)\citenamefont
		{O’Keeffe}, \citenamefont {Bolognesi}, \citenamefont {Coreno},
		\citenamefont {Moise}, \citenamefont {Richter}, \citenamefont {Cautero},
		\citenamefont {Stebel}, \citenamefont {Sergo}, \citenamefont {Pravica},
		\citenamefont {Ovcharenko},\ and\ \citenamefont {Avaldi}}]{Keeffe:2011}%
	\BibitemOpen
	\bibfield  {author} {\bibinfo {author} {\bibfnamefont {P.}~\bibnamefont
			{O’Keeffe}}, \bibinfo {author} {\bibfnamefont {P.}~\bibnamefont
			{Bolognesi}}, \bibinfo {author} {\bibfnamefont {M.}~\bibnamefont {Coreno}},
		\bibinfo {author} {\bibfnamefont {A.}~\bibnamefont {Moise}}, \bibinfo
		{author} {\bibfnamefont {R.}~\bibnamefont {Richter}}, \bibinfo {author}
		{\bibfnamefont {G.}~\bibnamefont {Cautero}}, \bibinfo {author} {\bibfnamefont
			{L.}~\bibnamefont {Stebel}}, \bibinfo {author} {\bibfnamefont
			{R.}~\bibnamefont {Sergo}}, \bibinfo {author} {\bibfnamefont
			{L.}~\bibnamefont {Pravica}}, \bibinfo {author} {\bibfnamefont
			{Y.}~\bibnamefont {Ovcharenko}}, \ and\ \bibinfo {author} {\bibfnamefont
			{L.}~\bibnamefont {Avaldi}},\ }\href@noop {} {\bibfield  {journal} {\bibinfo
			{journal} {Rev. Sci. Instrum.}\ }\textbf {\bibinfo {volume} {82}},\ \bibinfo
		{pages} {033109} (\bibinfo {year} {2011})}\BibitemShut {NoStop}%
	\bibitem [{\citenamefont {Rouz\'ee}\ \emph {et~al.}(2011)\citenamefont
		{Rouz\'ee}, \citenamefont {Johnsson}, \citenamefont {Gryzlova}, \citenamefont
		{Fukuzawa}, \citenamefont {Yamada}, \citenamefont {Siu}, \citenamefont
		{Huismans}, \citenamefont {Louis}, \citenamefont {Bijkerk}, \citenamefont
		{Holland}, \citenamefont {Grum-Grzhimailo}, \citenamefont {Kabachnik},
		\citenamefont {Vrakking},\ and\ \citenamefont {Ueda}}]{Rouzee:2011}%
	\BibitemOpen
	\bibfield  {author} {\bibinfo {author} {\bibfnamefont {A.}~\bibnamefont
			{Rouz\'ee}}, \bibinfo {author} {\bibfnamefont {P.}~\bibnamefont {Johnsson}},
		\bibinfo {author} {\bibfnamefont {E.~V.}\ \bibnamefont {Gryzlova}}, \bibinfo
		{author} {\bibfnamefont {H.}~\bibnamefont {Fukuzawa}}, \bibinfo {author}
		{\bibfnamefont {A.}~\bibnamefont {Yamada}}, \bibinfo {author} {\bibfnamefont
			{W.}~\bibnamefont {Siu}}, \bibinfo {author} {\bibfnamefont {Y.}~\bibnamefont
			{Huismans}}, \bibinfo {author} {\bibfnamefont {E.}~\bibnamefont {Louis}},
		\bibinfo {author} {\bibfnamefont {F.}~\bibnamefont {Bijkerk}}, \bibinfo
		{author} {\bibfnamefont {D.~M.~P.}\ \bibnamefont {Holland}}, \bibinfo
		{author} {\bibfnamefont {A.~N.}\ \bibnamefont {Grum-Grzhimailo}}, \bibinfo
		{author} {\bibfnamefont {N.~M.}\ \bibnamefont {Kabachnik}}, \bibinfo {author}
		{\bibfnamefont {M.~J.~J.}\ \bibnamefont {Vrakking}}, \ and\ \bibinfo {author}
		{\bibfnamefont {K.}~\bibnamefont {Ueda}},\ }\href@noop {} {\bibfield
		{journal} {\bibinfo  {journal} {Phys. Rev. A}\ }\textbf {\bibinfo {volume}
			{83}},\ \bibinfo {pages} {031401} (\bibinfo {year} {2011})}\BibitemShut
	{NoStop}%
	\bibitem [{\citenamefont {Lyamayev}\ \emph {et~al.}(2013)\citenamefont
		{Lyamayev}, \citenamefont {Ovcharenko}, \citenamefont {Katzy}, \citenamefont
		{Devetta}, \citenamefont {Bruder}, \citenamefont {LaForge}, \citenamefont
		{Mudrich}, \citenamefont {Person}, \citenamefont {Stienkemeier},
		\citenamefont {Krikunova}, \citenamefont {M{\"o}ller}, \citenamefont
		{Piseri}, \citenamefont {Avaldi}, \citenamefont {Coreno}, \citenamefont
		{O’Keeffe}, \citenamefont {Bolognesi}, \citenamefont {Alagia},
		\citenamefont {Kivim{\"a}ki}, \citenamefont {Fraia}, \citenamefont {Brauer},
		\citenamefont {Drabbels}, \citenamefont {Mazza}, \citenamefont {Stranges},
		\citenamefont {Finetti}, \citenamefont {Grazioli}, \citenamefont {Plekan},
		\citenamefont {Richter}, \citenamefont {Prince},\ and\ \citenamefont
		{Callegari}}]{Lyamayev:2013}%
	\BibitemOpen
	\bibfield  {author} {\bibinfo {author} {\bibfnamefont {V.}~\bibnamefont
			{Lyamayev}}, \bibinfo {author} {\bibfnamefont {Y.}~\bibnamefont
			{Ovcharenko}}, \bibinfo {author} {\bibfnamefont {R.}~\bibnamefont {Katzy}},
		\bibinfo {author} {\bibfnamefont {M.}~\bibnamefont {Devetta}}, \bibinfo
		{author} {\bibfnamefont {L.}~\bibnamefont {Bruder}}, \bibinfo {author}
		{\bibfnamefont {A.}~\bibnamefont {LaForge}}, \bibinfo {author} {\bibfnamefont
			{M.}~\bibnamefont {Mudrich}}, \bibinfo {author} {\bibfnamefont
			{U.}~\bibnamefont {Person}}, \bibinfo {author} {\bibfnamefont
			{F.}~\bibnamefont {Stienkemeier}}, \bibinfo {author} {\bibfnamefont
			{M.}~\bibnamefont {Krikunova}}, \bibinfo {author} {\bibfnamefont
			{T.}~\bibnamefont {M{\"o}ller}}, \bibinfo {author} {\bibfnamefont
			{P.}~\bibnamefont {Piseri}}, \bibinfo {author} {\bibfnamefont
			{L.}~\bibnamefont {Avaldi}}, \bibinfo {author} {\bibfnamefont
			{M.}~\bibnamefont {Coreno}}, \bibinfo {author} {\bibfnamefont
			{P.}~\bibnamefont {O’Keeffe}}, \bibinfo {author} {\bibfnamefont
			{P.}~\bibnamefont {Bolognesi}}, \bibinfo {author} {\bibfnamefont
			{M.}~\bibnamefont {Alagia}}, \bibinfo {author} {\bibfnamefont
			{A.}~\bibnamefont {Kivim{\"a}ki}}, \bibinfo {author} {\bibfnamefont {M.~D.}\
			\bibnamefont {Fraia}}, \bibinfo {author} {\bibfnamefont {N.~B.}\ \bibnamefont
			{Brauer}}, \bibinfo {author} {\bibfnamefont {M.}~\bibnamefont {Drabbels}},
		\bibinfo {author} {\bibfnamefont {T.}~\bibnamefont {Mazza}}, \bibinfo
		{author} {\bibfnamefont {S.}~\bibnamefont {Stranges}}, \bibinfo {author}
		{\bibfnamefont {P.}~\bibnamefont {Finetti}}, \bibinfo {author} {\bibfnamefont
			{C.}~\bibnamefont {Grazioli}}, \bibinfo {author} {\bibfnamefont
			{O.}~\bibnamefont {Plekan}}, \bibinfo {author} {\bibfnamefont
			{R.}~\bibnamefont {Richter}}, \bibinfo {author} {\bibfnamefont {K.~C.}\
			\bibnamefont {Prince}}, \ and\ \bibinfo {author} {\bibfnamefont
			{C.}~\bibnamefont {Callegari}},\ }\href
	{http://stacks.iop.org/0953-4075/46/i=16/a=164007} {\bibfield  {journal}
		{\bibinfo  {journal} {J. Phys. B}\ }\textbf {\bibinfo {volume} {46}},\
		\bibinfo {pages} {164007} (\bibinfo {year} {2013})}\BibitemShut {NoStop}%
	\bibitem [{\citenamefont {Fukuzawa}\ \emph {et~al.}(2016)\citenamefont
		{Fukuzawa}, \citenamefont {Tachibana}, \citenamefont {Motomura},
		\citenamefont {Xu}, \citenamefont {Nagaya}, \citenamefont {Wada},
		\citenamefont {Johnsson}, \citenamefont {Siano}, \citenamefont {Mondal},
		\citenamefont {Ito}, \citenamefont {Kimura}, \citenamefont {Sakai},
		\citenamefont {Matsunami}, \citenamefont {Hayashita}, \citenamefont
		{Kajikawa}, \citenamefont {Liu}, \citenamefont {Robert}, \citenamefont
		{Miron}, \citenamefont {Feifel}, \citenamefont {Marangos}, \citenamefont
		{Tono}, \citenamefont {Inubushi}, \citenamefont {Yabashi}, \citenamefont
		{Yao},\ and\ \citenamefont {Ueda}}]{Fukuzawa:2016}%
	\BibitemOpen
	\bibfield  {author} {\bibinfo {author} {\bibfnamefont {H.}~\bibnamefont
			{Fukuzawa}}, \bibinfo {author} {\bibfnamefont {T.}~\bibnamefont {Tachibana}},
		\bibinfo {author} {\bibfnamefont {K.}~\bibnamefont {Motomura}}, \bibinfo
		{author} {\bibfnamefont {W.~Q.}\ \bibnamefont {Xu}}, \bibinfo {author}
		{\bibfnamefont {K.}~\bibnamefont {Nagaya}}, \bibinfo {author} {\bibfnamefont
			{S.}~\bibnamefont {Wada}}, \bibinfo {author} {\bibfnamefont {P.}~\bibnamefont
			{Johnsson}}, \bibinfo {author} {\bibfnamefont {M.}~\bibnamefont {Siano}},
		\bibinfo {author} {\bibfnamefont {S.}~\bibnamefont {Mondal}}, \bibinfo
		{author} {\bibfnamefont {Y.}~\bibnamefont {Ito}}, \bibinfo {author}
		{\bibfnamefont {M.}~\bibnamefont {Kimura}}, \bibinfo {author} {\bibfnamefont
			{T.}~\bibnamefont {Sakai}}, \bibinfo {author} {\bibfnamefont
			{K.}~\bibnamefont {Matsunami}}, \bibinfo {author} {\bibfnamefont
			{H.}~\bibnamefont {Hayashita}}, \bibinfo {author} {\bibfnamefont
			{J.}~\bibnamefont {Kajikawa}}, \bibinfo {author} {\bibfnamefont {X.-J.}\
			\bibnamefont {Liu}}, \bibinfo {author} {\bibfnamefont {E.}~\bibnamefont
			{Robert}}, \bibinfo {author} {\bibfnamefont {C.}~\bibnamefont {Miron}},
		\bibinfo {author} {\bibfnamefont {R.}~\bibnamefont {Feifel}}, \bibinfo
		{author} {\bibfnamefont {J.~P.}\ \bibnamefont {Marangos}}, \bibinfo {author}
		{\bibfnamefont {K.}~\bibnamefont {Tono}}, \bibinfo {author} {\bibfnamefont
			{Y.}~\bibnamefont {Inubushi}}, \bibinfo {author} {\bibfnamefont
			{M.}~\bibnamefont {Yabashi}}, \bibinfo {author} {\bibfnamefont
			{M.}~\bibnamefont {Yao}}, \ and\ \bibinfo {author} {\bibfnamefont
			{K.}~\bibnamefont {Ueda}},\ }\href
	{http://stacks.iop.org/0953-4075/49/i=3/a=034004} {\bibfield  {journal}
		{\bibinfo  {journal} {J. Phys. B}\ }\textbf {\bibinfo {volume} {49}},\
		\bibinfo {pages} {034004} (\bibinfo {year} {2016})}\BibitemShut {NoStop}%
	\bibitem [{\citenamefont {Kornilov}\ \emph {et~al.}(2010)\citenamefont
		{Kornilov}, \citenamefont {Wang}, \citenamefont {B{\"u}nermann},
		\citenamefont {Healy}, \citenamefont {Leonard}, \citenamefont {Peng},
		\citenamefont {Leone}, \citenamefont {Neumark},\ and\ \citenamefont
		{Gessner}}]{Kornilov:2010}%
	\BibitemOpen
	\bibfield  {author} {\bibinfo {author} {\bibfnamefont {O.}~\bibnamefont
			{Kornilov}}, \bibinfo {author} {\bibfnamefont {C.~C.}\ \bibnamefont {Wang}},
		\bibinfo {author} {\bibfnamefont {O.}~\bibnamefont {B{\"u}nermann}}, \bibinfo
		{author} {\bibfnamefont {A.~T.}\ \bibnamefont {Healy}}, \bibinfo {author}
		{\bibfnamefont {M.}~\bibnamefont {Leonard}}, \bibinfo {author} {\bibfnamefont
			{C.}~\bibnamefont {Peng}}, \bibinfo {author} {\bibfnamefont {S.~R.}\
			\bibnamefont {Leone}}, \bibinfo {author} {\bibfnamefont {D.~M.}\ \bibnamefont
			{Neumark}}, \ and\ \bibinfo {author} {\bibfnamefont {O.}~\bibnamefont
			{Gessner}},\ }\href {\doibase 10.1021/jp907312t} {\bibfield  {journal}
		{\bibinfo  {journal} {J. Phys. Chem. A}\ }\textbf {\bibinfo {volume} {114}},\
		\bibinfo {pages} {1437} (\bibinfo {year} {2010})}\BibitemShut {NoStop}%
	\bibitem [{\citenamefont {Sch\"utte}\ \emph {et~al.}(2014)\citenamefont
		{Sch\"utte}, \citenamefont {Arbeiter}, \citenamefont {Fennel}, \citenamefont
		{Vrakking},\ and\ \citenamefont {Rouz\'ee}}]{Schuette:2014}%
	\BibitemOpen
	\bibfield  {author} {\bibinfo {author} {\bibfnamefont {B.}~\bibnamefont
			{Sch\"utte}}, \bibinfo {author} {\bibfnamefont {M.}~\bibnamefont {Arbeiter}},
		\bibinfo {author} {\bibfnamefont {T.}~\bibnamefont {Fennel}}, \bibinfo
		{author} {\bibfnamefont {M.~J.~J.}\ \bibnamefont {Vrakking}}, \ and\ \bibinfo
		{author} {\bibfnamefont {A.}~\bibnamefont {Rouz\'ee}},\ }\href {\doibase
		10.1103/PhysRevLett.112.073003} {\bibfield  {journal} {\bibinfo  {journal}
			{Phys. Rev. Lett.}\ }\textbf {\bibinfo {volume} {112}},\ \bibinfo {pages}
		{073003} (\bibinfo {year} {2014})}\BibitemShut {NoStop}%
	\bibitem [{\citenamefont {Kling}\ \emph {et~al.}(2014)\citenamefont {Kling},
		\citenamefont {Paul}, \citenamefont {Gura}, \citenamefont {Laurent},
		\citenamefont {De}, \citenamefont {Li}, \citenamefont {Wang}, \citenamefont
		{Ahn}, \citenamefont {Kim}, \citenamefont {Kim}, \citenamefont {Litvinyuk},
		\citenamefont {Cocke}, \citenamefont {Ben-Itzhak}, \citenamefont {Kim},\ and\
		\citenamefont {Kling}}]{Kling:2014}%
	\BibitemOpen
	\bibfield  {author} {\bibinfo {author} {\bibfnamefont {N.~G.}\ \bibnamefont
			{Kling}}, \bibinfo {author} {\bibfnamefont {D.}~\bibnamefont {Paul}},
		\bibinfo {author} {\bibfnamefont {A.}~\bibnamefont {Gura}}, \bibinfo {author}
		{\bibfnamefont {G.}~\bibnamefont {Laurent}}, \bibinfo {author} {\bibfnamefont
			{S.}~\bibnamefont {De}}, \bibinfo {author} {\bibfnamefont {H.}~\bibnamefont
			{Li}}, \bibinfo {author} {\bibfnamefont {Z.}~\bibnamefont {Wang}}, \bibinfo
		{author} {\bibfnamefont {B.}~\bibnamefont {Ahn}}, \bibinfo {author}
		{\bibfnamefont {C.~H.}\ \bibnamefont {Kim}}, \bibinfo {author} {\bibfnamefont
			{T.~K.}\ \bibnamefont {Kim}}, \bibinfo {author} {\bibfnamefont {I.~V.}\
			\bibnamefont {Litvinyuk}}, \bibinfo {author} {\bibfnamefont {C.~L.}\
			\bibnamefont {Cocke}}, \bibinfo {author} {\bibfnamefont {I.}~\bibnamefont
			{Ben-Itzhak}}, \bibinfo {author} {\bibfnamefont {D.}~\bibnamefont {Kim}}, \
		and\ \bibinfo {author} {\bibfnamefont {M.~F.}\ \bibnamefont {Kling}},\ }\href
	{http://stacks.iop.org/1748-0221/9/i=05/a=P05005} {\bibfield  {journal}
		{\bibinfo  {journal} {Journal of Instrumentation}\ }\textbf {\bibinfo
			{volume} {9}},\ \bibinfo {pages} {P05005} (\bibinfo {year}
		{2014})}\BibitemShut {NoStop}%
	\bibitem [{\citenamefont {Sch\"utte}\ \emph {et~al.}(2015)\citenamefont
		{Sch\"utte}, \citenamefont {Lahl}, \citenamefont {Oelze}, \citenamefont
		{Krikunova}, \citenamefont {Vrakking},\ and\ \citenamefont
		{Rouz\'ee}}]{Schuette:2015}%
	\BibitemOpen
	\bibfield  {author} {\bibinfo {author} {\bibfnamefont {B.}~\bibnamefont
			{Sch\"utte}}, \bibinfo {author} {\bibfnamefont {J.}~\bibnamefont {Lahl}},
		\bibinfo {author} {\bibfnamefont {T.}~\bibnamefont {Oelze}}, \bibinfo
		{author} {\bibfnamefont {M.}~\bibnamefont {Krikunova}}, \bibinfo {author}
		{\bibfnamefont {M.~J.~J.}\ \bibnamefont {Vrakking}}, \ and\ \bibinfo {author}
		{\bibfnamefont {A.}~\bibnamefont {Rouz\'ee}},\ }\href@noop {} {\bibfield
		{journal} {\bibinfo  {journal} {Phys. Rev. Lett.}\ }\textbf {\bibinfo
			{volume} {114}},\ \bibinfo {pages} {123002} (\bibinfo {year}
		{2015})}\BibitemShut {NoStop}%
	\bibitem [{\citenamefont {Eppink}\ and\ \citenamefont
		{Parker}(1997)}]{Eppink:1997}%
	\BibitemOpen
	\bibfield  {author} {\bibinfo {author} {\bibfnamefont {A.~T. J.~B.}\
			\bibnamefont {Eppink}}\ and\ \bibinfo {author} {\bibfnamefont {D.~H.}\
			\bibnamefont {Parker}},\ }\href@noop {} {\bibfield  {journal} {\bibinfo
			{journal} {Rev. Sci. Instrum.}\ }\textbf {\bibinfo {volume} {68}},\ \bibinfo
		{pages} {3477} (\bibinfo {year} {1997})}\BibitemShut {NoStop}%
	\bibitem [{\citenamefont {Offerhaus}\ \emph {et~al.}(2001)\citenamefont
		{Offerhaus}, \citenamefont {Nicole}, \citenamefont {Lépine}, \citenamefont
		{Bordas}, \citenamefont {Rosca-Pruna},\ and\ \citenamefont
		{Vrakking}}]{Offerhaus:2001}%
	\BibitemOpen
	\bibfield  {author} {\bibinfo {author} {\bibfnamefont {H.~L.}\ \bibnamefont
			{Offerhaus}}, \bibinfo {author} {\bibfnamefont {C.}~\bibnamefont {Nicole}},
		\bibinfo {author} {\bibfnamefont {F.}~\bibnamefont {Lépine}}, \bibinfo
		{author} {\bibfnamefont {C.}~\bibnamefont {Bordas}}, \bibinfo {author}
		{\bibfnamefont {F.}~\bibnamefont {Rosca-Pruna}}, \ and\ \bibinfo {author}
		{\bibfnamefont {M.~J.~J.}\ \bibnamefont {Vrakking}},\ }\href@noop {}
	{\bibfield  {journal} {\bibinfo  {journal} {Rev. Sci. Instrum.}\ }\textbf
		{\bibinfo {volume} {72}},\ \bibinfo {pages} {3245} (\bibinfo {year}
		{2001})}\BibitemShut {NoStop}%
	\bibitem [{\citenamefont {Skruszewicz}\ \emph {et~al.}(2014)\citenamefont
		{Skruszewicz}, \citenamefont {Passig}, \citenamefont {Przystawik},
		\citenamefont {Truong}, \citenamefont {K{\"o}ther}, \citenamefont
		{Tiggesb{\"a}umker},\ and\ \citenamefont {Meiwes-Broer}}]{Skruszewicz:2014}%
	\BibitemOpen
	\bibfield  {author} {\bibinfo {author} {\bibfnamefont {S.}~\bibnamefont
			{Skruszewicz}}, \bibinfo {author} {\bibfnamefont {J.}~\bibnamefont {Passig}},
		\bibinfo {author} {\bibfnamefont {A.}~\bibnamefont {Przystawik}}, \bibinfo
		{author} {\bibfnamefont {N.}~\bibnamefont {Truong}}, \bibinfo {author}
		{\bibfnamefont {M.}~\bibnamefont {K{\"o}ther}}, \bibinfo {author}
		{\bibfnamefont {J.}~\bibnamefont {Tiggesb{\"a}umker}}, \ and\ \bibinfo
		{author} {\bibfnamefont {K.-H.}\ \bibnamefont {Meiwes-Broer}},\ }\href@noop
	{} {\bibfield  {journal} {\bibinfo  {journal} {Int. J. Mass Spectrom.}\
		}\textbf {\bibinfo {volume} {365–366}},\ \bibinfo {pages} {338 } (\bibinfo
		{year} {2014})}\BibitemShut {NoStop}%
	\bibitem [{\citenamefont {Krishnan}\ \emph {et~al.}(2011)\citenamefont
		{Krishnan}, \citenamefont {Fechner}, \citenamefont {Kremer}, \citenamefont
		{Sharma}, \citenamefont {Fischer}, \citenamefont {Camus}, \citenamefont
		{Jha}, \citenamefont {Krishnamurthy}, \citenamefont {Pfeifer}, \citenamefont
		{Moshammer} \emph {et~al.}}]{KrishnanPRL:2011}%
	\BibitemOpen
	\bibfield  {author} {\bibinfo {author} {\bibfnamefont {S.}~\bibnamefont
			{Krishnan}}, \bibinfo {author} {\bibfnamefont {L.}~\bibnamefont {Fechner}},
		\bibinfo {author} {\bibfnamefont {M.}~\bibnamefont {Kremer}}, \bibinfo
		{author} {\bibfnamefont {V.}~\bibnamefont {Sharma}}, \bibinfo {author}
		{\bibfnamefont {B.}~\bibnamefont {Fischer}}, \bibinfo {author} {\bibfnamefont
			{N.}~\bibnamefont {Camus}}, \bibinfo {author} {\bibfnamefont
			{J.}~\bibnamefont {Jha}}, \bibinfo {author} {\bibfnamefont {M.}~\bibnamefont
			{Krishnamurthy}}, \bibinfo {author} {\bibfnamefont {T.}~\bibnamefont
			{Pfeifer}}, \bibinfo {author} {\bibfnamefont {R.}~\bibnamefont {Moshammer}},
		\emph {et~al.},\ }\href@noop {} {\bibfield  {journal} {\bibinfo  {journal}
			{Phys. Rev. Lett.}\ }\textbf {\bibinfo {volume} {107}},\ \bibinfo {pages}
		{173402} (\bibinfo {year} {2011})}\BibitemShut {NoStop}%
	\bibitem [{\citenamefont {Krishnan}\ \emph {et~al.}(2012)\citenamefont
		{Krishnan}, \citenamefont {Peltz}, \citenamefont {Fechner}, \citenamefont
		{Sharma}, \citenamefont {Kremer}, \citenamefont {Fischer}, \citenamefont
		{Camus}, \citenamefont {Pfeifer}, \citenamefont {Jha}, \citenamefont
		{Krishnamurthy}, \citenamefont {Schr{\"o}ter}, \citenamefont {Ullrich},
		\citenamefont {Stienkemeier}, \citenamefont {Moshammer}, \citenamefont
		{Fennel},\ and\ \citenamefont {Mudrich}}]{Krishnan:2012}%
	\BibitemOpen
	\bibfield  {author} {\bibinfo {author} {\bibfnamefont {S.~R.}\ \bibnamefont
			{Krishnan}}, \bibinfo {author} {\bibfnamefont {C.}~\bibnamefont {Peltz}},
		\bibinfo {author} {\bibfnamefont {L.}~\bibnamefont {Fechner}}, \bibinfo
		{author} {\bibfnamefont {V.}~\bibnamefont {Sharma}}, \bibinfo {author}
		{\bibfnamefont {M.}~\bibnamefont {Kremer}}, \bibinfo {author} {\bibfnamefont
			{B.}~\bibnamefont {Fischer}}, \bibinfo {author} {\bibfnamefont
			{N.}~\bibnamefont {Camus}}, \bibinfo {author} {\bibfnamefont
			{T.}~\bibnamefont {Pfeifer}}, \bibinfo {author} {\bibfnamefont
			{J.}~\bibnamefont {Jha}}, \bibinfo {author} {\bibfnamefont {M.}~\bibnamefont
			{Krishnamurthy}}, \bibinfo {author} {\bibfnamefont {C.-D.}\ \bibnamefont
			{Schr{\"o}ter}}, \bibinfo {author} {\bibfnamefont {J.}~\bibnamefont
			{Ullrich}}, \bibinfo {author} {\bibfnamefont {F.}~\bibnamefont
			{Stienkemeier}}, \bibinfo {author} {\bibfnamefont {R.}~\bibnamefont
			{Moshammer}}, \bibinfo {author} {\bibfnamefont {T.}~\bibnamefont {Fennel}}, \
		and\ \bibinfo {author} {\bibfnamefont {M.}~\bibnamefont {Mudrich}},\
	}\href@noop {} {\bibfield  {journal} {\bibinfo  {journal} {New J. Phys.}\
		}\textbf {\bibinfo {volume} {14}},\ \bibinfo {pages} {075016} (\bibinfo
		{year} {2012})}\BibitemShut {NoStop}%
	\bibitem [{\citenamefont {Heidenreich}\ \emph {et~al.}(2016)\citenamefont
		{Heidenreich}, \citenamefont {Grüner}, \citenamefont {Rometsch},
		\citenamefont {Krishnan}, \citenamefont {Stienkemeier},\ and\ \citenamefont
		{Mudrich}}]{Heidenreich:2016}%
	\BibitemOpen
	\bibfield  {author} {\bibinfo {author} {\bibfnamefont {A.}~\bibnamefont
			{Heidenreich}}, \bibinfo {author} {\bibfnamefont {B.}~\bibnamefont
			{Grüner}}, \bibinfo {author} {\bibfnamefont {M.}~\bibnamefont {Rometsch}},
		\bibinfo {author} {\bibfnamefont {S.~R.}\ \bibnamefont {Krishnan}}, \bibinfo
		{author} {\bibfnamefont {F.}~\bibnamefont {Stienkemeier}}, \ and\ \bibinfo
		{author} {\bibfnamefont {M.}~\bibnamefont {Mudrich}},\ }\href
	{http://stacks.iop.org/1367-2630/18/i=7/a=073046} {\bibfield  {journal}
		{\bibinfo  {journal} {New J. Phys.}\ }\textbf {\bibinfo {volume} {18}},\
		\bibinfo {pages} {073046} (\bibinfo {year} {2016})}\BibitemShut {NoStop}%
	\bibitem [{\citenamefont {Heidenreich}\ \emph {et~al.}(2017)\citenamefont
		{Heidenreich}, \citenamefont {Grüner}, \citenamefont {Schomas},
		\citenamefont {Stienkemeier}, \citenamefont {Krishnan},\ and\ \citenamefont
		{Mudrich}}]{Heidenreich:2017}%
	\BibitemOpen
	\bibfield  {author} {\bibinfo {author} {\bibfnamefont {A.}~\bibnamefont
			{Heidenreich}}, \bibinfo {author} {\bibfnamefont {B.}~\bibnamefont
			{Grüner}}, \bibinfo {author} {\bibfnamefont {D.}~\bibnamefont {Schomas}},
		\bibinfo {author} {\bibfnamefont {F.}~\bibnamefont {Stienkemeier}}, \bibinfo
		{author} {\bibfnamefont {S.~R.}\ \bibnamefont {Krishnan}}, \ and\ \bibinfo
		{author} {\bibfnamefont {M.}~\bibnamefont {Mudrich}},\ }\href@noop {}
	{\bibfield  {journal} {\bibinfo  {journal} {J. Mod. Opt.}\ }\textbf {\bibinfo
			{volume} {64}},\ \bibinfo {pages} {1061} (\bibinfo {year}
		{2017})}\BibitemShut {NoStop}%
	\bibitem [{\citenamefont {Wituschek}\ \emph {et~al.}(2016)\citenamefont
		{Wituschek}, \citenamefont {von Vangerow}, \citenamefont {Grzesiak},
		\citenamefont {Stienkemeier},\ and\ \citenamefont
		{Mudrich}}]{Wituschek:2016}%
	\BibitemOpen
	\bibfield  {author} {\bibinfo {author} {\bibfnamefont {A.}~\bibnamefont
			{Wituschek}}, \bibinfo {author} {\bibfnamefont {J.}~\bibnamefont {von
				Vangerow}}, \bibinfo {author} {\bibfnamefont {J.}~\bibnamefont {Grzesiak}},
		\bibinfo {author} {\bibfnamefont {F.}~\bibnamefont {Stienkemeier}}, \ and\
		\bibinfo {author} {\bibfnamefont {M.}~\bibnamefont {Mudrich}},\ }\href@noop
	{} {\bibfield  {journal} {\bibinfo  {journal} {Rev. Sci. Instrum.}\ }\textbf
		{\bibinfo {volume} {87}},\ \bibinfo {pages} {083105} (\bibinfo {year}
		{2016})}\BibitemShut {NoStop}%
\end{thebibliography}

%

\end{document}